\documentclass[11pt]{article}
\usepackage{amssymb,amsthm}
\usepackage{amsmath}
\usepackage{amscd}
\usepackage{latexsym}

\def\beq{\begin{equation}}
\def\eeq{\end{equation}}
\def\bea{\begin{eqnarray}}
\def\eea{\end{eqnarray}}

\catcode`@=11
\renewcommand{\section}{\@startsection{section}{1}{0pt}{\medskipamount}
{\medskipamount}{\large\bf}}
\numberwithin{equation}{section}
\catcode`@=12
\def\a{\alpha}
\def\b{\beta}
\def\g{\gamma}

\def\de{\delta}
\def\ve{\varepsilon}
\def\h{\eta}

\def\la{\lambda}
\def\m{\mu}

\def\n{\nu}
\def\r{\rho}

\def\s{\sigma}
\def\p{\phi}
\def\vp{\varphi}

\def\c{\chi}
\def\ome{\omega}

\def\La{\Lambda}

\def\Xt{\tilde X}
\def\et{\tilde e}

\def\1{{\bar 1}}
\def\2{{\bar 2}}
\def\3{{\bar 3}}
\def\4{{\bar 4}}

\def\lrc{\,\lrcorner\,}
\def\hra{\,\hookrightarrow\,}

\newcommand{\Xh}{\hat{X}}

\newcommand{\eh}{\hat{e}}

\newcommand{\zb}{\bar{z}}

\newcommand{\su}{{{\rm SU}(2)}}

\newcommand{\uo}{{{\rm U}(1)}}
\newcommand{\urmL}{{{\mathfrak u}}}

\newcommand{\urm}{{{\rm U}}}
\newcommand{\surm}{{{\rm SU}}}
\newcommand{\surmL}{{{\mathfrak{su}}}}
\newcommand{\sorm}{{{\rm SO}}}
\newcommand{\sormL}{{{\mathfrak{so}}}}

\newcommand{\gfrak}{{\mathfrak{g}}}
\newcommand{\hfrak}{{\mathfrak{h}}}

\newcommand{\Hom}{{\rm Hom}}
\newcommand{\End}{{\rm End}}

\newcommand{\one}{{\bf 1}}
\newcommand{\diag}{{\rm diag}}
\newcommand{\C}{\mathbb C}
\newcommand{\R}{\mathbb R}

\newcommand{\Z}{\mathbb Z}

\newcommand{\Acal}{{\cal A}}

\newcommand{\Fcal}{{\cal F}}
\newcommand{\Ecal}{{\cal E}}
\newcommand{\Lcal}{{\cal L}}

\def\im{{\rm i}}
\def\e{{\,\rm e}\,}
\def\N2{$N{=}2$}
\def\pa{\partial}
\def\diff{{\rm d}}
\def\tr{{\rm tr}}
\def\sfrac#1#2{{\textstyle\frac{#1}{#2}}}
\def\>{\rangle}
\def\<{\langle}
\def\+{\dagger}
\def\={\ =\ }
\def\und{\qquad\textrm{and}\qquad}
\def\and{\qquad\textrm{and}\qquad}
\def\for{\qquad\textrm{for}\quad}
\def\with{\qquad\textrm{with}\quad}

\begin{document}
\begin{titlepage}
\setcounter{page}{0}
\begin{flushright}
EMPG--13--15\\
ITP--UH--18/13
\end{flushright}

\vskip 2.5cm

\begin{center}

{\Large\bf Orbifold instantons, moment maps\\[10pt] and
  Yang-Mills theory with sources
}

\vspace{15mm}

{\large Tatiana A. Ivanova}${}^1$, \
{\large Olaf Lechtenfeld}${}^2$, \ {\large Alexander D. Popov${}^{2}$}
 \ and  \ {\large Richard J. Szabo${}^3$}
\\[5mm]
\noindent ${}^1${\em Bogoliubov Laboratory of Theoretical Physics, JINR} \\
{\em  141980 Dubna, Moscow Region, Russia}\\
Email: {\tt ita@theor.jinr.ru}
\\[3mm]
\noindent ${}^2${\em Institut f\"ur Theoretische Physik} and
{\em Riemann Center for Geometry and Physics}\\
{\em Leibniz Universit\"at Hannover}\\
{\em Appelstra\ss e 2, 30167 Hannover, Germany}\\
Email: {\tt lechtenf@itp.uni-hannover.de, popov@itp.uni-hannover.de}
\\[3mm]
\noindent ${}^3${\em Department of Mathematics, Heriot-Watt University\\
Colin Maclaurin Building, Riccarton, Edinburgh EH14 4AS, U.K.}\\
and
{\em Maxwell Institute
  for Mathematical Sciences, Edinburgh, U.K.}\\
and
{\em The Tait Institute, Edinburgh, U.K.}\\
{Email: {\tt R.J.Szabo@hw.ac.uk}}

\vspace{15mm}

\begin{abstract}
\noindent
We revisit the problem of constructing instantons on ADE orbifolds
$\R^4/\Gamma$ and point out some subtle relations with the complex
structure on the orbifold. We consider generalized instanton equations on
$\R^4/\Gamma$ which are BPS equations for the Yang-Mills equations
with an external current. The relation between level sets
of the moment maps in the hyper-K\"ahler quotient construction of the
instanton moduli space and sources in the Yang-Mills
equations is discussed. We describe two types of spherically-symmetric
$\Gamma$-equivariant connections on complex V-bundles
 over $\R^4/\Gamma$ which are tailored to the way in which the orbifold group
 acts on the fibres. Some explicit abelian and nonabelian $\su$-invariant solutions
to the instanton equations on the orbifold are worked out.
\end{abstract}

\end{center}

\end{titlepage}

\section{Introduction and summary}

\noindent
Instantons in Yang-Mills theory~\cite{1} and gravity~\cite{2,3} play an important
role in modern field theory~\cite{4,5,6}. They are nonperturbative
configurations which solve
first order (anti-)self-duality equations for the gauge field and the
Riemann curvature tensor, respectively.
The construction of gauge instantons can be described systematically in the framework of twistor theory~\cite{7,8} and by the ADHM construction~\cite{9}. There are also many methods for constructing
gravitational instantons including twistor theory~\cite{8} and the hyper-K\"ahler quotient
construction~\cite{10} based on the hyper-K\"ahler moment map introduced in~\cite{11}.

In this paper we revisit the problem of constructing instantons on the
ADE orbifolds $\R^4/\Gamma$. The corresponding instanton moduli spaces
are of special interest in type~II string theory, where they can be realized as Higgs
branches of certain quiver gauge theories which appear as worldvolume
field theories on D$p$-branes in a D$p$-D$(p{+}4)$ system with the
D$(p{+}4)$-branes located at the fixed point of the
orbifold~\cite{DM}. The ADHM equations can be identified with the
vacuum equations of the supersymmetric gauge theory, and the structure
of the vacuum moduli space provides an important example of resolution
of spacetime singularities by stringy effects in the form of D-brane probes. We point out in particular some salient relations
between the construction of instantons and complex structures on
$\R^4/\Gamma$.

Kronheimer~\cite{10} considers $\Gamma$-equivariant solutions
of the matrix equations
\beq\label{1.1}\begin{array}{l}
[W_2,W_3]+[W_1,W_4]=\Xi_1\ ,\\[4pt]
[W_3,W_1]+[W_2,W_4]=\Xi_2\ ,\\[4pt]
[W_1,W_2]+[W_3,W_4]=\Xi_3\ ,
\end{array}
\eeq
where $\Gamma$ is a finite subgroup of the Lie group $\su$ acting on
the fundamental representation
$\C^2\cong\R^4$, $W_\m$ with $\m=1,2,3,4$ are matrices taking values
in e.g.\footnote{In type~II string theory in the presence of
  orientifold O$(p{+}4)$ planes one should use instead the Lie algebras
  of orthogonal
  or symplectic Lie groups.} the Lie algebra
$\urmL(N)$, and $\Xi_a$ with $a=1,2,3$ are matrices in the center $\hfrak$ of a subalgebra
$\gfrak$ of $\urmL(N)$. For $\Xi_a=0$ the equations (\ref{1.1}) are  the anti-self-dual Yang-Mills
equations on the orbifold $\C^2/\Gamma$ reduced by translations. Their solutions satisfy the
full Yang-Mills equations. In the general case, the equations (\ref{1.1}) are interpreted as hyper-K\"ahler moment map
quotient equations, and Hitchin shows~\cite{12} that one can similarly interpret
the Bogomolny monopole equations and vortex equations. Kronheimer
shows that the moduli space of
solutions to (\ref{1.1}) in the Coulomb branch is a hyper-K\"ahler ALE
space $M_\xi$, which is the minimal resolution
\beq\label{1.2}
M_\xi \ \longrightarrow \ M_0
\eeq
of the orbifold $M_0=\C^2/\Gamma$. Here $\xi$ are parameters in the matrices $\Xi_a$ of (\ref{1.1}).
Similar results were obtained in~\cite{13,14} for $\su$-invariant Yang-Mills instantons on $\R^4$
(see also~\cite{15}). Moreover, it was shown
by Kronheimer and Nakajima~\cite{KN} that there exists a bundle
$\Ecal\to M_{\xi}$ with Chern classes $c_1(\Ecal )=0$
and $c_2(\Ecal )=(\#\Gamma-1)/\#\Gamma$ such that the moduli space of
framed instantons on $\Ecal$ satisfying the anti-self-dual
Yang-Mills equations coincides with the base
manifold $M_\xi$ itself. In the limit $\xi =0$ one obtains
$\C^2/\Gamma$ as the moduli space of minimal fractional instantons on the V-bundle $\Ecal$ over the orbifold
$M_0= \C^2/\Gamma$.

In this paper we consider gauge instanton equations with matrices
$\Xi_a$ on the orbifold
$\C^2/\Gamma\cong\R^4/\Gamma$
and show that the choices of $\Xi_a\ne 0$ correspond to sources in the Yang-Mills equations.
For gauge potentials on $\R^4/\Gamma$
with $\Gamma =\Z_{k+1}$ we analyse solutions of $\Gamma$-equivariance conditions in two different
$\su$-invariant bases adapted to the spherical symmetry. Recall that one can write a realization
of the Lie algebra $\sormL(4)\cong \surmL(2)\oplus \surmL(2)$ in terms
of vector fields on $\R^4/\Gamma$ as
\beq\label{1.3}
E^a=-\eta^a_{\m\n}\,y^\m\, \frac{\pa}{\pa y^\n}\und \tilde E^a=-\bar\eta^a_{\m\n}\,
y^\m\, \frac{\pa}{\pa y^\n}\ ,
\eeq
where $\eta^a_{\m\n}$ and $\bar\eta^a_{\m\n}$ are components of the
self-dual and anti-self-dual 't~Hooft
tensors~\cite{16} and $y^\m$ are local coordinates on $\R^4/\Gamma$. The commutation relations between these vector fields are
\beq\label{1.4}
[E^a,E^b]=2\ve^{abc}\, E^c\ ,\qquad [\tilde E^a,\tilde
E^b]=2\ve^{abc}\, \tilde E^c \and [E^a,\tilde E^b]=0\ .
\eeq
Introducing complex coordinates $z^1=y^1+\im\, y^2$ and $z^2=y^3+\im\, y^4$ on $\R^4/\Gamma\cong\C^2/\Gamma$,
one finds that the vector fields $\tilde E^a$ preserve
this complex structure but the vector fields $E^a$ do not, i.e.~the
group $\su$ acting on $\C^2$ is
generated by $(\tilde E^a)$. Furthermore, the actions of the
corresponding Lie
derivatives are given by
\beq\label{1.5}
\Lcal_{\tilde E^b} e^a=0\und \Lcal_{\tilde E^b} \et^a
=2\ve^a_{bc}\, \et^c \ ,
\eeq
where $e^a=e^a_\m\, \diff y^\m$ and $\et^a=\et^a_\m\, \diff y^\m$ are one-forms dual to the vector fields
$E^a$ and $\tilde E^a$, respectively. We show that both bases of
one-forms $(e^a,\diff r)$ and $(\tilde
e^a,\diff r)$ with $r^2=\de_{\m\n}\, y^\m\, y^\n$ can be used for describing spherically-symmetric
instanton configurations, but due to (\ref{1.5}) the basis $(e^a, \diff r)$ is more suitable
for connections on V-bundles $\Ecal$ with trivial action of the finite group $\Gamma\subset\su$,
while the basis $(\et^a, \diff r)$ is more suitable for connections on V-bundles $\Ecal$
with non-trivial $\Gamma$-action on the fibres
of $\Ecal$. Explicit examples of abelian and nonabelian $\su$-invariant
instanton solutions on $\R^4/\Z_{k+1}$ are worked out below.

The structure of the remainder of this paper is as follows. In
Section~2 we consider generalized instanton equations on $\R^4$ which
reduce to (\ref{1.1}) and show that they correspond to BPS-type
equations for Yang-Mills theory with sources. In Section~3 we extend
these equations to the ADE quotient singularities $\R^4/\Gamma$,
focusing on the special case $\Gamma=\Z_{k+1}$. In Section~4 we study
the moduli spaces of translationally-invariant instantons on
$\R^4/\Gamma$ via the
hyper-K\"ahler quotient construction. In Section~5 we consider the
construction of spherically-symmetric instanton solutions on
$\R^4/\Gamma$ and make some preliminary comments concerning the structure of the
instanton moduli spaces, though a detailed description of these moduli
spaces is beyond the scope of the present work.

\bigskip

\section{Instanton equations on $\R^4$}

\noindent {\bf Euclidean space $\R^4$. \ }
Consider the two-forms \beq
\label{2.1} \ome^a:=\sfrac12\, \eta^a_{\m\n}\, \diff y^\m\wedge\diff
y^\n\ , \eeq where $y^\m$ are coordinates on $\R^4$ and
$\ome^a_{\m\n}:=\eta^a_{\m\n}$ are components of the 't~Hooft
tensors given by the formulas \beq \label{2.2}
\eta^a_{bc}=\ve^a_{bc}\and \eta^a_{b4}=-\eta^a_{4b}=\de^a_b\ .
\eeq
Here $\ve^1_{23}=1$, $\m ,\n ,\ldots=1,2,3,4$ and $a,b,\ldots =1,2,3$. The forms
$\ome^a$ are symplectic and self-dual,
\beq \label{2.3}
\diff\ome^a=0\und\ast\ome^a=\ome^a\ , \eeq where $\ast$ is the
Hodge duality operator for the flat metric
\beq \label{2.4}
g=\de_{\m\n}\, \diff y^\m\otimes \diff y^\n
\eeq
on $\R^4$.

Using the metric (\ref{2.4}) we introduce three complex structures $J^a=\ome^a\circ g^{-1}$ on $\R^4$ with components
\beq \label{2.5}
(J^a)^\m_\n=\ome^a_{\n\la}\, \de^{\la\m}\ ,
\eeq
so that $(\R^4, J^a)\cong\C^2_{J^a}$. The space $\R^4$ is
hyper-K\"ahler, i.e.~it is
K\"ahler with respect to each of the complex structures (\ref{2.5}).
We choose one of them, $J^3=:J$, to identify $\R^4$ and $\C^2\cong (\R^4, J)$.
With respect to $J$ the complex two-form
\beq\label{2.6}
\ome_{\C}=\ome^1 + \im\, \ome^2
\eeq
is closed and holomorphic, i.e.~$\ome_{\C}$ is a $(2,0)$-form.

\bigskip

\noindent {\bf Instanton equations. \ } Let $\Ecal$ be a rank $N$
complex vector bundle over $\R^4\cong\C^2$. We endow this bundle
with a connection $\Acal=\Acal_\mu\, \diff y^\mu$ of curvature $\Fcal =\diff\Acal
+\Acal\wedge\Acal=\frac12\,\Fcal_{\mu\nu}\, \diff y^\mu\wedge \diff y^\nu$ taking values in the Lie algebra $\urmL(N)$. Let us
constrain the curvature $\Fcal$ by the equations
\beq\label{2.7}
*\Fcal + \Fcal = 2\ome^a\, \Xi_a\ , \eeq
where the functions $\Xi_a$
belong to $\urmL(N)$. Solutions to this equation of finite topological
charge are called (generalized) instantons. If $\Xi_a$ belong to the center
$\urmL(1)$ of $\urmL(N)$ and $\diff\Xi_a=0$, then solutions to the equations
(\ref{2.7}) satisfy the Yang-Mills equations on $\R^4$. If $\Xi_a$ do not
belong to this center,\footnote{Later on we will consider an
important example of such non-central elements $\Xi_a$.} then (\ref{2.7}) are BPS-type
equations for Yang-Mills theory with {\it sources} which vanish
only if $\Xi_a$ are constant and $\Xi_a\in \urmL(1)\subset \urmL(N)$ for
$a=1,2,3$. Indeed, from (\ref{2.7}) we get \beq\label{2.8}
\diff*\Fcal + \Acal\wedge *\Fcal - *\Fcal\wedge\Acal =
2\ome^a\wedge\big(\diff\Xi_a+[\Acal ,\Xi_a] \big) \ , \eeq which after
taking the Hodge dual can be rewritten as \beq\label{2.9}
\pa_\m\Fcal_{\m\n} + [\Acal_\m, \Fcal_{\m\n}]=
4\ome^a_{\m\n}\, \big(\pa_\m\Xi_a+[\Acal_\m ,\Xi_a] \big) \ . \eeq The current \beq\label{2.10} j_\m := 4\ome^a_{\n\m}\, D_\n\Xi_a \with
D_\m\Xi_a: =\pa_\m\Xi_a+[\Acal_\m ,\Xi_a] \eeq satisfies the covariant continuity
equation \beq\label{2.11} D_\m j_\m =0\ , \eeq as required for minimal coupling of an external
current in the Yang-Mills equations.

\bigskip

\noindent {\bf Variational equations. \ } To formulate the generalized
instanton equations (\ref{2.7}) as absolute minima of Euler-Lagrange
equations derived from an action priniciple, we note that the presence of the current
(\ref{2.10}) in the Yang-Mills equations (\ref{2.9}) requires the
addition of the
term \beq\label{2.12} \sfrac12\,\tr\, j_\m\, \Acal_{\m} \eeq in the
standard Yang-Mills lagrangian \beq\label{2.13} \Lcal_{\sf
{YM}}=-\sfrac18\, \tr\, \Fcal_{\m\n}\, \Fcal_{\m\n}\ . \eeq Up to a total
derivative the term (\ref{2.12}) is equivalent to the term
\beq\label{2.14} \sfrac12\,\ome^a_{\m\n}\,\tr\, \Fcal_{\m\n}\, \Xi_a\
. \eeq After adding the term (\ref{2.14}), together with the non-dynamical term
\beq\label{2.15} -3\,\tr\, \Xi_a\, \Xi_a \eeq and the topological
density \beq\label{2.16} -\sfrac12\,\ve_{\m\n\la\s}\,\tr\,
\Fcal_{\m\n}\, \Fcal_{\la\s}\ , \eeq we obtain the lagrangian
\beq\label{2.17} \Lcal = -\sfrac14\,\tr\, \big(\Fcal^+_{\m\n}-
\ome^a_{\m\n}\, \Xi_a\big)\, \big(\Fcal^+_{\m\n}- \ome^a_{\m\n}\,
\Xi_a \big)\ , \eeq
where \beq\label{2.18}
\Fcal^+=\sfrac12\,(*\Fcal+\Fcal) \eeq is the
self-dual part of the curvature two-form $\Fcal$. In the following we will consider
constant matrices $\Xi_a$ for which (\ref{2.15}) becomes constant
and the term (\ref{2.14}) is topological. Constant
matrices of the form $\Xi_a=\im\,\xi_a\, \one_N$ correspond to D3-branes
in a
non-zero $B$-field in
string theory and can be described in terms of a noncommutative
deformation of Yang-Mills theory on the space $\R^4$ (see e.g.~\cite{17,18}).

\bigskip

\section{Instanton equations on $\R^4/\Gamma$}

\noindent {\bf Orbifold $\R^4/\Gamma$. \ } The complex structure
$J=J^3$, introduced in (\ref{2.5}), defines the complex
coor\-dinates \beq\label{3.1} z^1=y^1+\im\,y^2\und
z^2=y^3+\im\,y^4 \eeq on $\R^4\cong\C^2$, where $y^\m$ are real
coordinates. The Lie group $\su$ naturally acts on the vector space
$\C^2$ with the coordinates (\ref{3.1}). We are interested in the Kleinian
orbifolds $\C^2/\Gamma$ where $\Gamma$ is a finite subgroup of $\su$.
They have an ADE classification in which $\Gamma$ is associated
with the extended Dynkin
diagram of a simply-laced simple Lie algebra. For the $A_k$-type simple singularities,
corresponding to the cyclic group $\Gamma=\Z_{k+1}$ of order
$k{+}1$, explicit descriptions of instantons will be readily available. However,
most of our results can be generalized to the other ADE groups
$\Gamma$ corresponding to nonabelian orbifolds $\C^2/\Gamma$.

The action of $\Gamma=\Z_{k+1}$ on
$\C^2$ is given by
\beq\label{3.2} (z^1, z^2) \ \longmapsto \
\big(\zeta\, z^1, \zeta^{-1}\, z^2 \big)\ , \eeq where \beq\label{3.3}
\zeta=\exp\big(\mbox{$\frac{2\pi\,\im}{k+1}$}\big
) \with \zeta^{k+1}=1\eeq is a primitive $(k{+}1)$-th root of unity. This action has a single
isolated fixed point at the origin $(z^1, z^2)=(0,0)$.
The orbifold $\C^2/\Gamma$ is defined as the set of equivalence
classes on $\C^2$ with respect to the equivalence relation \beq\label{3.4}
\big(\zeta\, z^1, \zeta^{-1}\, z^2\big)\equiv (z^1, z^2)\ , \eeq
and it has a singularity at the origin.
The metric on
$\C^2/\Gamma$ is \beq\label{3.5} g=\diff z^1\otimes \diff \bar z^\1 +\diff
z^2 \otimes \diff \bar z^\2\ , \eeq where the coordinates $\bar z^\1 , \bar z^\2$ are
complex conjugated to $z^1, z^2$.

\bigskip

\noindent {\bf V-bundles on $\C^2/\Gamma$. \ } A V-bundle on
$\C^2/\Gamma$ is a $\Gamma$-equivariant bundle over $\C^2$, i.e.~a
vector bundle on $\C^2$ with a $\Gamma$-action on the
fibres which is compatible with the action of $\Gamma$ on $\C^2$. The orbifold group
$\Gamma=\Z_{k+1}$ has $k{+}1$ one-dimensional irreducible
representations such that the generator of $\Z_{k+1}$ acts on the
$\ell$-th $\Gamma$-module as multiplication by $\zeta^{\ell}$ for
$\ell=0,1,\ldots,k$. Let us denote by
$\Ecal_\ell$ complex V-bundles over $\C^2/\Gamma$ of rank $N_\ell$
on which $\Gamma$ acts in the $\ell$-th irreducible representation as
\beq\label{3.6}
v_\ell \ \longmapsto \ \zeta^\ell\, v_\ell \for
v_\ell\in\C^{N_\ell} \eeq
on a generic fibre $\C^{N_\ell}$ of
$\Ecal_\ell$.
Then
any complex V-bundle $\Ecal$ over
$\C^2/\Gamma$ of rank $N$ can be
decomposed into isotopical components as a Whitney sum \beq \label{3.7} \Ecal
=\bigoplus^k_{\ell=0}\, \Ecal_{\ell} \ , \eeq and its structure group
is of the form
\beq \label{3.8}
\prod\limits^k_{\ell=0}\, \urm(N_{\ell})\with
\sum\limits^k_{\ell=0}\, N_{\ell}=N\ . \eeq
From (\ref{3.6}) it
follows that the action of the point group on the V-bundle (\ref{3.7})
is given by the unitary
matrices
\beq \label{3.12}
v \ \longmapsto \ \g_{_\Gamma}(v) \with \g_{_\Gamma} =
\mathop{\bigoplus}\limits^k_{\ell=0}\, \zeta^{\ell}\,\one_{N_\ell}
\eeq
on vectors $v=(v_\ell)_{\ell=0}^k$ in the generic fibre
$\C^N=\bigoplus_{\ell=0}^k\, \C^{N_\ell}$ of $\Ecal$.

Simplifying the
situation discussed in the previous section, we choose
matrices $\Xi_a$ in the form
\beq \label{3.9}
\Xi_a
=\mathop{\bigoplus}\limits^k_{\ell=0}\, \im\;\xi_a^{\ell}\;\one_{N_{\ell}}\
, \eeq where $\xi_a^{\ell}\in \R$ are constants. The
matrices (\ref{3.9}) belong to the center of the Lie algebra of
the gauge group (\ref{3.8}). The diagonal $\uo$ subgroup of scalars in
(\ref{3.8}) acts trivially on $(\Ecal ,\Acal)$, so we can factor the gauge group (\ref{3.8}) by this
$\uo$ subgroup to get the quotient group \beq \label{3.10} G:=\Big
(\, \prod\limits^k_{\ell=0}\, \urm(N_{\ell})\, \Big)\, \big/\, \uo  \
. \eeq
Then the Lie algebra $\gfrak$ of $G$ is the traceless part of the Lie algebra of (\ref{3.8}), and one
should impose on $\xi_a^{\ell}$ in (\ref{3.9}) the tracelessness condition
\beq\label{3.11} \sum\limits^k_{\ell=0}\, \xi^{\ell}_a \, N_{\ell}=0\ , \eeq
which defines the center $\hfrak$ of $\gfrak$.

\bigskip

\noindent
{\bf $\Gamma$-equivariant connections. \ } Consider a one-form
\beq \label{3.13}
W=W_\m\, \diff y^\m=W_{z^1}\,\diff z^1 + W_{z^2}\,\diff z^2 + W_{\zb^\1}\,\diff \zb^\1+
W_{\zb^\2}\,\diff\zb^\2
\eeq
on $\R^4\cong\C^2$ which is invariant under the action of $\Gamma \subset\su\subset\sorm(4)$
defined by (\ref{3.2}). Then on the components
\beq \label{3.14}
W_{z^1}=\sfrac12\, (W_1-\im \, W_2) \and W_{z^2}=\sfrac12\, (W_3-\im\, W_4)
\eeq
the action of $\Gamma$ is given by
\beq \label{3.15}
W_{z^1} \ \longmapsto \ \zeta^{-1}\, W_{z^1}\and W_{z^2} \ \longmapsto
\ \zeta \, W_{z^2}\ .
\eeq

The action of $\Gamma$ on the components $\Acal_\m$ of any unitary connection
$\Acal =\Acal_\m\, \diff y^\m$ on
a hermitian V-bundle
(\ref{3.7}) is given by a combination of the spacetime action
(\ref{3.15}) and the adjoint action generated by (\ref{3.12}) as
\beq \label{3.16}
\Acal_{z^1}\ \longmapsto \ \zeta^{-1}\, \g_{_\Gamma}\, \Acal_{z^1}\, \g_{_\Gamma}^{-1}\and
\Acal_{z^2} \ \longmapsto \ \zeta\, \g_{_\Gamma}\, \Acal_{z^2}\,
\g_{_\Gamma}^{-1} \ .
\eeq
The corresponding $\Gamma$-equivariance conditions require that the
connection defines a covariant representation of the orbifold group, in the sense that
\beq \label{3.17}
\g_{_\Gamma}\, \Acal_{z^1}\, \g_{_\Gamma}^{-1}=\zeta\, \Acal_{z^1}\and
\g_{_\Gamma}\, \Acal_{z^2}\, \g_{_\Gamma}^{-1}=\zeta^{-1}\, \Acal_{z^2}\ .
\eeq
It is easy to see that the solutions to the constraint equations
(\ref{3.17}) are given by block off-diagonal matrices
\beq \label{3.18}
\Acal_{z^1}=\begin{pmatrix}
   0&0&\cdots&0&\psi_{k+1}\\\psi_1&0&\ddots&0&0\\0&\psi_2&\ddots&\vdots&0\\\vdots&\ddots&\ddots&0&0\\
   0&\cdots&0&\psi_k&0  \end{pmatrix}\and
\Acal_{z^2}=\begin{pmatrix}
   0&\phi_1&0&\cdots&0\\0&0&\phi_2&\ddots&\vdots\\
   \vdots&\vdots&\ddots&\ddots&0\\0&0& \cdots &0&\phi_k \\
   \phi_{k+1}&0& \cdots &0&0 \end{pmatrix}
\eeq
together with $\Acal_{\zb^\1}=-\Acal^\+_{z^1}$ and $ \Acal_{\zb^\2}=-\Acal^\+_{z^2}$.
Here the bundle morphisms $\psi_{\ell+1}:\Ecal_\ell\to\Ecal_{\ell+1}$
and $\phi_{\ell+1}:\Ecal_{\ell+1}\to\Ecal_\ell$ are bifundamental
scalar fields given fibrewise by matrices
\beq
\psi_{\ell+1} \ \in \ \Hom\big(\C^{N_\ell}\,,\,\C^{N_{\ell+1}}\big) \and
\phi_{\ell+1} \ \in \ \Hom\big(\C^{N_{\ell+1}}\,,\,\C^{N_\ell}\big)
\eeq
for $\ell=0,1,\dots,k$
(with indices read modulo~$k{+}1$). Substitution of
(\ref{3.18}) in (\ref{2.7}) then yields the generalized instanton
equations on the orbifold $\C^2/\Gamma$.
The transformations (\ref{3.15}) are defined for
the holonomic basis
$\diff y^\m$ of one-forms on $\R^4/\Gamma$ and can differ for other bases of one-forms, leading to
modifications of the formulas (\ref{3.16})--(\ref{3.18}).

\bigskip

\section{Translationally-invariant instantons}

\noindent
{\bf Matrix equations. \ } Consider translationally-invariant connections $\Acal$ on the V-bundle (\ref{3.7})
over $\C^2/\Gamma$ satisfying the equations (\ref{2.7}) with $\Xi_a$
given in (\ref{3.9}), i.e.~we assume that $\Acal_\m$ are independent of the coordinates $y^\m$,
which reduces (\ref{2.7}) to the matrix equations
(\ref{1.1}) with $W_\m :=\Acal_\m$. Denoting $B_1:=\Acal_{z^1}$ and $B_2:=\Acal_{z^2}$ for
$\Acal$ given by (\ref{3.18}) with constant matrices $\psi_{\ell +1}$ and
$\phi_{\ell +1}$ for $\ell=0,1,\dots,k$, we obtain the equations
\beq \label{4.1}
\sfrac12\,\h^a_{\m\n}\, [W_\m , W_\n]=\Xi_a\ ,
\eeq
which can be rewritten as
\bea \label{4.2}
[B_1, B_2]\=-\sfrac{\im}{4}\, (\Xi_1 -\im\, \Xi_2) &=:& \Xi_{\C}\ , \\[4pt]
 \label{4.3}
\big[B_1, B_1^\+\big]+\big[B_2, B_2^\+\big]\=-\sfrac{\im}{2}\, \Xi_3
&=:& \Xi_{\R}\ .
\eea
Solutions to these equations satisfy the reduced Yang-Mills equations (\ref{2.9})
with the external source
\beq \label{4.4}
j_\m=-4\h^a_{\m\n}\, [W_\n , \Xi_a]\ ,
\eeq
where $W_\m$ is given by (\ref{3.18}) and $\Xi_a$ by (\ref{3.9}).

\bigskip

\noindent{\bf Hyper-K\"ahler quotients. \ } The reduced equations
(\ref{4.1}) (and also the instanton equations (\ref{2.7}))
can be interpreted as hyper-K\"ahler moment map equations. For this, recall that if $(M,g,\ome^a)$ is a
hyper-K\"ahler manifold with an action of a Lie group $G$ which
preserves the metric $g$ and the three
K\"ahler forms~\footnote{They are K\"ahler with respect to the three complex structures $J^a=\ome^a\circ g^{-1}$.
With respect to the complex structure $J^3$, the two-form $\omega_\R= \ome^3$ is K\"ahler and $\ome_{\C}=
\ome^1+\im\,\ome^2$ is holomorphic.} $\ome^a$, then one can define three moment maps
\beq \label{4.5}
\m^a\, : \, M \ \longrightarrow \ \gfrak^*
\eeq
taking values in the dual $\gfrak^*$ of the Lie algebra $\gfrak$ of $G$ such that, for each
$\xi\in \gfrak$ with triholomorphic Killing vector field
$L_\xi$ generated by the $G$-action on $M$, the functions (\ref{4.5}) satisfy the equations
\beq \label{4.6}
\langle\diff\m^a, \xi \rangle=L_\xi\lrc\,\ome^a\ ,
\eeq
where $\langle-,-\rangle$ is the dual pairing between elements of
$\gfrak^*$ and $\gfrak$, and $\lrc$ denotes contraction of vector
fields and differential forms. Denoting by $\m =(\m^1, \m^2, \m^3)$
the vector-valued moment map
\beq \label{4.7}
\m \, : \, M \ \longrightarrow \ \R^3\otimes\gfrak^*\ ,
\eeq
we can consider the $G$-invariant level set
\beq \label{4.8}
\m^{-1}(\Xi)
\eeq
which defines a submanifold of the manifold $M$, where $\Xi=(\Xi_1, \Xi_2, \Xi_3)\in\R^3\otimes\hfrak^*$ and $\hfrak$ is the center of
$\gfrak$.
Then one can define the hyper-K\"ahler quotient as (see e.g.~\cite{10,11,12})
\beq \label{4.9}
M_\xi =\m^{-1}(\Xi)\, \big/\!\!\big/\!\!\big/ \, G\ ,
\eeq
where $\xi=(\xi^\ell_a)$
are parameters defining $\Xi=(\Xi_a) \in \R^3\otimes\hfrak^*$. The
hyper-K\"ahler metric on $M$ descends to a hyper-K\"ahler metric on
the quotient $M_\xi$. When the group action is free, the reduced space
$M_\xi$ is a hyper-K\"ahler manifold of dimension $\dim
M_\xi = \dim M- 4\dim G$.

In the case of the matrix model (\ref{4.1}), the manifold $M$ is the flat hyper-K\"ahler manifold
\beq \label{4.10}
M=\R^4\otimes \urmL(N)\ ,
\eeq
the group $G$ is given in (\ref{3.10}) and the three moment maps are~\footnote{We identify $\urmL(N)^*$ and $\urmL(N)$.}
\beq \label{4.11}
\m^a(W)=\sfrac12\,\h^a_{\m\n}\, [W_\m , W_\n ] \ \in \ \urmL(N)\ .
\eeq
Solutions of the equations (\ref{4.2})--(\ref{4.3}) form a submanifold $\m^{-1}(\Xi)$ of
the manifold (\ref{4.10}), and by factoring with the gauge group
(\ref{3.10}) (which for generic parameters $\xi=(\xi^\ell_a)$ acts
freely on the solutions) we obtain the
moduli space (\ref{4.9}). This moduli space was studied by Kronheimer~\cite{10}, who showed
that for $\Gamma =\Z_{k+1}$ and the Coulomb branch $N_0=N_1=\cdots=N_k=1$ it is
a smooth
four-dimensional asymptotically locally euclidean (ALE) hyper-K\"ahler manifold $M_\xi$ with metric
defined by the parameters $\xi=(\xi^\ell_a)$. The ALE condition
means that at asymptotic infinity of $M_\xi$ the metric
approximates the euclidean metric on the orbifold
$\C^2/\Gamma$. Kronheimer also shows that $M_\xi$ is diffeomorphic to the
minimal smooth resolution of the Kleinian singularity $M_0=\C^2/\Gamma$,
regarded as the affine algebraic variety $x^{k+1}+y^2+z^2=0$ in $\C^3$. For
the Hilbert-Chow map
\beq \label{4.12}
\pi \,:\, M_\xi \ \longrightarrow \ M_0
\eeq
the exceptional divisor of the blow-up is the set
\beq \label{4.13}
\pi^{-1}(0)=\bigcup^k_{\ell =0}\, \Sigma_{\ell}\ ,
\eeq
where $\Sigma_{\ell}\cong\C P^1$ and $k=\#\Gamma -1$.\footnote{Recall that we consider $\Gamma =\Z_{k+1}$ for
definiteness here, but many of these considerations generalize to the
other Kleinian groups $\Gamma\subset\su$. In the general case,
$N_\ell$ are the dimensions of the irreducible representations of the
finite group $\Gamma$ in Kronheimer's construction.} The parameters
$\xi$ determine the periods of the three symplectic forms $\omega^a$
as
\beq
\int_{\Sigma_\ell}\, \omega^a = \xi^\ell_a \ .
\eeq
In the general case $N_\ell\geq0$, one can also define a map $M_\xi\to
M_0$ which is a resolution of singularities~\cite{Nakajima}.

\bigskip

\noindent
{\bf Hermitian Yang-Mills
  connections. \ } The matrix $\Xi_\C$ in (\ref{4.2}) parametrizes deformations of the complex
structure on the V-bundle $\Ecal$ and it can be reabsorbed through a non-analytic change of coordinates on the space
(\ref{4.10})~\cite{10, 19}. Therefore we may take $\Xi_\C = 0$ without
loss of generality; in this case the ALE space $M_\xi$ is
biholomorphic to the minimal resolution.
In fact, the moduli spaces $M_{\xi}$ and $M_{\xi'}$ are diffeomorphic for distinct $\xi$ and $\xi'$
such that $\Xi_\R \ne 0$ for both sets of parameters. For $\Xi_\C = 0$ we have
$\Xi_1=\Xi_2= 0$ and the equations (\ref{2.7}) become the hermitian Yang-Mills equations~\cite{20,21}
\beq \label{4.14}
*\Fcal +\Fcal=\ome^3\, \Xi_3\ .
\eeq
A connection $\Acal$ on $\Ecal$ satisfying (\ref{4.14}) is said to be
a hermitian Yang-Mills connection. It defines a
holomorphic structure on $\Ecal$ since from (\ref{4.14}) it follows
that the curvature $\Fcal$ is of type $(1,1)$
with respect to the complex structure $J$, i.e.
\beq \label{4.15}
\Fcal^{2,0}=0=\Fcal^{0,2}\ ,
\eeq
and the third equation from (\ref{4.14}),
\beq \label{4.16}
\ome^3_{\m\n}\, \Fcal_{\m\n}=\Xi_3\ ,
\eeq
means that for $\Xi_3=\im\,\xi\,\one_N$ the V-bundle $\Ecal$ is (semi-)stable~\cite{20,21}.
In the special case $\Xi_3=0$ we get the standard anti-self-dual Yang-Mills equations
\beq \label{4.17}
*\Fcal = -\Fcal\ .
\eeq

\bigskip

\noindent
{\bf Translationally-equivariant instantons. \ } Instead of constant
matrices $\Acal_\m$ which reduce (\ref{2.7}) to
the matrix equations (\ref{4.1}), one can also consider the gauge potential
\beq \label{4.18}
\Acal =\sfrac12\,\ome^a_{\m\n}\,\Xi_a\,y^\m\, \diff y^\n\ ,
\eeq
where the commuting matrices $\Xi_a$ are given in (\ref{3.9}). The
connection (\ref{4.18}) is translationally-invariant up to a gauge
transformation and can be extended to the orbifold $T^4/\Gamma$,
where $T^4$ is a four-dimensional torus.
The curvature of $\Acal$ is
\beq \label{4.19}
\Fcal=\diff\Acal =\sfrac12\,\ome^a_{\m\n}\,\Xi_a\,\diff
y^\m\wedge\diff y^\n \ ,
\eeq
providing in essence the three symplectic structures $\ome^a$ from (\ref{2.1}).

\bigskip

\section{Spherically-symmetric instantons}

\noindent
{\bf Cone $C(S^3/\Gamma)$. \ } The euclidean space $\R^4$ can be regarded as a cone over the three-sphere
$S^3$,
\beq \label{5.1}
\R^4\setminus \{0\}=C(S^3)
\eeq
with the metric
\beq \label{5.2}
g=\delta_{\m\n}\, \diff y^\m\otimes \diff y^\n = \diff r^2 + r^2\,
\delta_{ab} \, e^a\otimes e^b\ ,
\eeq
where $r^2=\delta_{\m\n} \, y^\m\, y^\n$ and $(e^a)$ give a basis of
left $\su$-invariant one-forms on $S^3$. One can define $e^a$
by the formula
\beq \label{5.3}
e^a:=-\frac{1}{r^2}\,\eta^a_{\m\n}\, y^{\m}\, \diff y^{\n} \ ,
\eeq
where the 't~Hooft tensors $\eta^a_{\m\n}$ are defined in (\ref{2.2}). The one-forms
$e^a$ are dual to the vector fields $E^a$ from (\ref{1.3}). By using the identities
\bea \label{5.4}
\ve^a_{bc}\, \h^b_{\m\n}\, \h^c_{\la\sigma}&=&\de_{\m\la}\, \h^a_{\n\sigma}-
\de_{\m\sigma}\, \h^a_{\n\la}-\de_{\n\la}\, \h^a_{\m\sigma}+
\de_{\n\sigma}\, \h^a_{\m\la}\ , \\[4pt] \label{5.5}
\de_{ab}\, \h^a_{\m\n}\, \h^b_{\la\sigma}&=&\de_{\m\la}\, \de_{\n\sigma}-
\de_{\m\sigma}\, \de_{\n\la}+\ve_{\m\n\la\sigma}\ ,
\eea
one can easily verify the Maurer-Cartan equations
\beq\label{5.6}
\diff e^a + \ve^a_{bc}\, e^b\wedge e^c =0 \eeq
and
\beq \label{5.7}
\ome^a=\sfrac12\,\h^a_{\m\n}\, \diff y^\m\wedge \diff
y^\n=\sfrac12\,\h^a_{\m\n}\, \hat e^\m\wedge
\hat e^\n\ , \eeq
where
\beq \label{5.8}
\hat e^a:=r \, e^a\and \hat e^4:=\diff r\ .
\eeq
The relation  (\ref{5.2}) between the metric in cartesian and spherical coordinates can be
readily checked as well.

All formulas (\ref{5.2})--(\ref{5.8}) are also valid for the orbifold $\C^2/\Gamma$ after imposing
the equivalence relation (\ref{3.4}), and the orbifold is a cone over
the lens space $S^3/\Gamma$,\footnote{The orbifolds $S^3/\Gamma$ for
  arbitrary ADE point groups $\Gamma$ exhaust the possible
  Sasaki-Einstein manifolds in three dimensions.}
\beq \label{5.9}
\big(\C^2\setminus \{0\}\big)\,\big/\, \Gamma = C\big(S^3/\Gamma \big)
\ ,
\eeq
with the metric (\ref{5.2}). The one-forms (\ref{5.3}) in the complex coordinates
(\ref{3.1}) have the form
\beq \label{5.10}
e^1+\im\,e^2=\sfrac{\im}{r^2}\, \big(z^1\, \diff z^2 - z^2\, \diff
z^1\big) \and
e^3+\im\,e^4=\sfrac{\im}{r^2}\, \big(\zb^\1\, \diff z^1 + \zb^\2\, \diff z^2\big)\ ,
\eeq
plus their complex conjugated expressions. Hence (\ref{5.10}) defines two complex one-forms which
are $(1,0)$-forms with respect to the complex structure $J=J^3$ defined in (\ref{2.5}). The symplectic
two-forms (\ref{5.7}) and the complex structures (\ref{2.5}) have the {\it same}
components in the holonomic $(\diff y^\m, \frac\pa{\pa y^\m})$ and non-holonomic $(\hat e^a, \hat E_a)$ bases,
where $\hat E_a\lrc\,\hat e^b=\de^b_a$. From (\ref{1.5}), (\ref{3.2}) and (\ref{5.10}) it follows that
\beq \label{5.11}
e^a\and e^4:=\frac{\diff r}{r}=\diff\tau \with \tau =\log r
\eeq
are invariant under the action of the finite group $\Gamma \subset\su$.

\bigskip

\noindent
{\bf Nahm equations. \ } Consider the complex V-bundle $\Ecal$
over $\C^2/\Gamma$ described in Section~3. Let
\beq \label{5.12}
\Acal =\Xh_\m\, \eh^\m=\sfrac12\,\big(\Xh_1-\im\,\Xh_2 \big)\,
\big(\eh^1 + \im\,\eh^2 \big)+
\sfrac12\,\big(\Xh_3-\im\,\Xh_4\big)\, \big(\eh^3 + \im\,\eh^4 \big) + \mbox{h.c.}
\eeq
be a connection on $\Ecal$ written in the basis (\ref{5.8}). The corresponding
$\Gamma$-equivariance conditions are
\beq \label{5.13}
\g_{_\Gamma}\, \big(\Xh_1-\im\,\Xh_2\big)\,
\g_{_\Gamma}^{-1}=\Xh_1-\im\,\Xh_2 \and
\g_{_\Gamma}\, \big(\Xh_3-\im\,\Xh_4\big)\, \g_{_\Gamma}^{-1}=\Xh_3-\im\,\Xh_4\ .
\eeq
Solutions to these equations are given by
\beq \label{5.14}\begin{array}{c}
\sfrac12\,\big(\Xh_1-\im\,\Xh_2\big)=\diag (\c_0,\c_1,\ldots,\c_k)\and
\sfrac12\,\big(\Xh_3-\im\,\Xh_4\big)=\diag (\vp_0,\vp_1,\ldots,\vp_k)\
, \\[4pt]
\sfrac12\,\big(\Xh_1+\im\,\Xh_2\big)=-\diag
\big(\c_0^\+,\c_1^\+,\ldots,\c_k^\+ \big)\and
\sfrac12\,\big(\Xh_3+\im\,\Xh_4\big)=-\diag
\big(\vp_0^\+,\vp_1^\+,\ldots,\vp_k^\+ \big)\ ,
\end{array}
\eeq
where $\c_\ell$ and $\vp_\ell$ are $N_\ell\times N_\ell$ complex matrices. Thus the
$\Gamma$-equivariance conditions in the basis (\ref{5.8}) forces the block-diagonal form (\ref{5.14})
of the connection components $\Xh_\m$, i.e.~the connection $\Acal$ is
reducible or else $N_\ell =0$ for $\ell\ne 0$ if
$\Gamma$ acts trivially on $\Ecal$.

The instanton equations  (\ref{2.7}) are conformally invariant and it is more convenient to
consider them on the cylinder
\beq \label{5.15}
\R\times S^3/\Gamma
\eeq
with the metric
\beq\label{5.16}
g_{\rm cyl} =\diff \tau^2 + \de_{ab}\, e^a \otimes e^b=
\frac{\diff  r^2}{r^2} + \de_{ab}\, e^a \otimes e^b=
\frac{1}{r^2}\, g\ .\eeq
In the basis $(e^\m )=(e^a, \diff\tau )$ the $\su$-invariant (spherically-symmetric) connection
$\Acal$ and its curvature $\Fcal$ have components depending only on
$r=\e^\tau$ and are given by
\beq\label{5.17}
\Acal = X_\m \, e^\m \with X_\m={r}\, \Xh_\m\ ,
\eeq
 \beq \label{5.18}
\Fcal_{4 a}= \frac{\diff X_a}{\diff \tau} + [X_\tau , X_a]\and \Fcal_{ab}= -2\ve_{abc}\,X_c + [X_a , X_b]\ ,
\eeq
and (\ref{2.7}) reduce to a form of the generalized Nahm equations
given by
\beq \label{5.19}
\frac{\diff X_a}{\diff \tau}=- [X_4 , X_a]-2X_a +
\frac{1}{2}\,\ve_{abc} \, [X_b , X_c]-\Xi_a\ .
\eeq
Introducing
\beq \label{5.20}
Y_\m:=\e^{2\tau}\, X_\m\and  s=\e^{-2\tau}=\sfrac{1}{r^2}\ ,
\eeq
we obtain the equations
\beq \label{5.21}
2\, \frac{\diff Y_a}{\diff s}= [Y_4 , Y_a]- \frac{1}{2}\,\ve_{abc} \,
[Y_b , Y_c]+\frac{1}{s^2}\, \Xi_a\ .
\eeq
For $\Xi_a=0$ these equations coincide with the Nahm equations~\cite{22}. Choosing  $\Xi_a=0$
and defining
\beq \label{5.22}
\alpha :=\sfrac12\, (Y_3+\im\,Y_4 )\and\b:=\sfrac12\, (Y_1+\im\,Y_2 )\ ,
\eeq
we obtain the equations
\bea \label{5.23}
\frac{\diff}{\diff s}\big(\a + \a^\+ \big) + \big[\a , \a^\+ \big]+
\big[\b , \b^\+ \big]&=& 0 \ ,
\\[4pt] \label{5.24}
\frac{\diff\b}{\diff s} + [\a , \b]&=& 0
\eea
considered by Kronheimer~\cite{13,14} (see also~\cite{15}) in the
description of $\su$-invariant instantons. The equations (\ref{5.21})
have three obvious solutions which we now consider in turn.

\bigskip

\noindent
{\bf Abelian instantons with $\Xi_a\ne 0$. \ }
For the first solution, we choose
\beq \label{5.25}
Y_a=-\sfrac{1}{2s}\,\Xi_a= -\sfrac{r^2}{2}\,\Xi_a\and Y_4=0\ .
\eeq
Then we get the solution
\beq \label{5.26}
\Acal = - \sfrac12\,r^2\, e^a\,\Xi_a\and \Fcal =
\sfrac12\,\h^a_{\m\n}\, \Xi_a\, \eh^\m\wedge\eh^\n
\eeq
of the equations (\ref{2.7}), which coincide with (\ref{4.18}) and (\ref{4.19}). This configuration
can also be regarded as a translationally-equivariant solution of the self-dual Yang-Mills equations
\beq \label{5.27}
*\Fcal =\Fcal\ ,
\eeq
i.e.~as an anti-instanton on $\R^4/\Gamma$ or $T^4/\Gamma$.

\bigskip

\noindent
{\bf Abelian instantons with poles. \ } For the second solution, considered in~\cite{14}, we put $\Xi_a=0$ and $\frac{\diff Y_a}{\diff s} =0$. Then the constant matrices $Y_\m$ satisfy the
reduced anti-self-dual Yang-Mills equations
\beq \label{5.28}
[Y_a, Y_4] + \sfrac12 \,\ve_{abc} \, [Y_b, Y_c]=0
\eeq
considered in~\cite{10} and discussed in Section~4. Solutions to (\ref{5.28})
are necessarily given by commuting matrices~\cite{10}, and one can choose them in the form
\beq \label{5.29}
Y_a= 2\Lambda^2\, \hat\Xi_a \and Y_4=0\ ,
\eeq
where $\hat\Xi_a$ have the form (\ref{3.9}) and $\Lambda$ is a scale parameter.
 For the corresponding gauge potential and its field strength, we obtain
\beq \label{5.30}
\Acal =X_a\, e^a = \frac{2\Lambda^2}{r^3}\,\hat\Xi_a\, \eh^a
\and
\Fcal = -\frac{2\Lambda^2}{r^4}\,\bar\eta^a_{\m\n}\,\hat\Xi_a\,
\eh^\m\wedge\eh^\n \ ,
\eeq
where $\eh^\m$ are given in (\ref{5.8}) and $\bar\eta^a_{\m\n}$ are the anti-self-dual 't~Hooft tensors
defined by
\beq\label{5.38}
\bar\eta^a_{bc}=\ve^a_{bc} \and \bar\eta^a_{b4}=-\bar\eta^a_{4b}=-\de^a_b
\ .
\eeq
Thus we obtain singular
abelian solutions with delta-function sources
in the Maxwell equations, as discussed by~\cite{14}. The gauge
potential $\Acal$ from (\ref{5.30}) can be regarded as an
asymptotic approximation of a smooth solution. Note also that
\beq \label{5.32}
\bar\ome^a:=-\frac{2\Lambda^2}{r^4}\,\bar\eta^a_{\m\n}\,\eh^\m\wedge\eh^\n
\eeq
can be viewed as three additional anti-self-dual symplectic forms
on the cone $(\R^4\setminus\{0\})/\Gamma =
C(S^3/\Gamma )$, complimentary to those given in (\ref{2.1}).

\bigskip

\noindent
{\bf 't~Hooft instantons on $\C^2/\Gamma$. \ }
For the third solution we choose $Y_4=Y_{\tau}=0=\Xi_a$
to get
\beq \label{5.33}
Y_a=\frac{2}{s+\Lambda^{-2}}\,I_a=\frac{2\Lambda^{2}\, r^2}{r^2+\Lambda^{2}}\,I_a
\with \Lambda\in \R\and [I_a, I_b] =\ve^c_{ab}\, I_c\ .
\eeq
Then for the anti-self-dual connection and curvature we obtain
\beq \label{5.34}
\Acal =\frac{2\Lambda^{2}}{r^2+\Lambda^{2}}\,e^a\, I_a\and
\Fcal =-\frac{2\Lambda^{2}}{\big(r^2+\Lambda^{2} \big)^2}\,\bar\eta^a_{\m\n}\,
I_a\, \hat e^\m\wedge\hat e^\n   \ ,
\eeq
where we used the relation $s=r^{-2}$. Here $I_a$ are the generators of the group $\su$ embedded in the
broken gauge group (\ref{3.10}), i.e.~there are $k{+}1$ instanton solutions with gauge group
$\su\subset\urm(N_\ell)$ if $N_\ell\ge 2$ for all $\ell=0,1,\ldots,k$. From the explicit form of $e^a$ in
(\ref{5.3}) it follows that each of these solutions is the standard
't~Hooft instanton generalized from $\R^4$ to
$\R^4/\Gamma$. For framed instantons~\footnote{Framed instantons are
  instanton solutions modulo $\su$-invariant gauge
transformations which approach the identity at asymptotic infinity.}
there are four moduli: the scale
parameter $\Lambda$ and three global $\su$ rotational parameters (see e.g.~\cite{19}).

\bigskip

\noindent
{\bf Moduli spaces of $\boldsymbol{\su}$-invariant instantons. \ }
In the special case where $\Gamma$ is the trivial group, we obtain
$\su$-invariant solutions of the anti-self-dual Yang-Mills equations (\ref{4.17})
on $\R^4\setminus \{0\}=C(S^3)$.
The moduli spaces of these framed instantons (subject to
appropriate boundary conditions)
are four-dimensional hyper-K\"ahler ALE spaces $M_{\xi}$ resolving $M_0=\C^2/\Gamma'$
as in (\ref{4.12}), where $\Gamma^\prime$ is a finite subgroup of the group $\su$ related to
boundary conditions for the solutions~\cite{13}--\cite{15}. This is
the moduli space of the spherically-symmetric instanton which has the
minimal topological charge $c_2(\Ecal)=(\#\Gamma'-1)/\#\Gamma'$. In our
reducible case we obtain a product of
hyper-K\"ahler moduli spaces
\beq \label{5.35}
M_{\xi_0}\times M_{\xi_1}\times\cdots\times M_{\xi_k}\ .
\eeq
Note that $M_{\xi_\ell}$ is a point if $N_\ell =1$. For $N_\ell =1$ one can also use the
singular abelian solution from (\ref{5.30}),
\beq \label{5.36}
\Fcal^\ell =-\frac{2\Lambda^{2}}{r^4}\,\bar\eta^a_{\m\n}\,\xi_a^\ell\,\eh^\m\wedge\eh^\n   \ ,
\eeq
with $\xi_a^\ell\in\R$.

We have seen that for
constant matrices $Y_a,Y_\tau$ the moduli space is the orbifold
$M_0=\C^2/\Gamma$. For $s$-dependent
solutions $Y_a,Y_\tau$, similarly to~\cite{10,15} one can choose
boundary conditions such that each block tends to a constant multiple
of the identity $\one_{N_\ell}$ in the limits $\tau\to\pm\, \infty$,
while as $\tau\to0$ the solutions define a representation of $\su$.
For $N_0=N_1=\cdots=N_k=1$ it is natural to expect that the
corresponding moduli space of solutions is a resolution of the
orbifold $\C^2/\Gamma$.

\bigskip

\noindent
{\bf BPST instantons on $\C^2/\Gamma$. \ }
Instead of the one-forms (\ref{5.3}),
one can introduce a basis of right $\su$-invariant one-forms on $S^3/\Gamma$ given by
\beq\label{5.37}
\tilde e^a:= -
\frac{1}{r^2}\ \bar\eta^a_{\m\n}\, y^{\m}\, \diff y^{\n} \ . \eeq
They are dual to the vector fields $\tilde E^a$ given in (\ref{1.3}),
and they satisfy the relations
\beq\label{5.39}
\diff \tilde e^a +\ve^a_{bc}\,\tilde e^b\wedge \tilde e^c = 0\ ,
\eeq
\beq\label{5.40}
g =\delta_{\m\n}\, \diff y^{\m} \otimes \diff y^{\n}=\diff r^2 + r^2\,
\de_{ab}\, \tilde e^a\otimes \tilde e^b
\eeq
which are similar to those for $e^a$ and can be proven by using
identities for $\bar\eta^a_{\m\n}$ analogous to (\ref{5.4})--(\ref{5.5}).

The complex combinations
\beq\label{5.41}
\tilde e^1 +\im\, \tilde e^2 = \sfrac{\im}{r^2}\, \big(z^1\, \diff \zb^\2
- \zb^\2\, \diff z^1 \big)\and
\tilde e^3 +\im\, \tilde e^4 = \sfrac{\im}{r^2}\, \big(\zb^\1\, \diff z^1
+ z^2\, \diff \zb^\2 \big)
\eeq
are neither $(1,0)$- nor $(0,1)$-forms with respect to the complex
structure $J=J^3$. One can show that the forms
(\ref{5.41}) are $(1,0)$-forms with respect to the complex structure
\beq\label{5.42}
\tilde J=\tilde J^3:=\big(\bar\eta^3_{\m\la}\, \de^{\la\n} \big)
\eeq
which is used in the consideration of self-duality equations (and anti-instantons) on $\R^4/\Gamma$.
The one-forms (\ref{5.10}) and (\ref{5.41}) are related by the
coordinate change $z^2\mapsto\zb^\2$ or,
equivalently, by the change of orientation $x^4\mapsto -x^4$ of $\R^4/\Gamma$. For a fixed orientation,
this inequivalence becomes more apparent in the case of the $\C P^2$, $K3$ and ALE
hyper-K\"ahler manifolds. Note that exactly $\tilde e^a$ (but not $e^a$) form a basis of one-forms
on the Sasaki-Einstein space $S^3/\Gamma\subset \R^4/\Gamma$, since the complex structure on $\C P^1\hra S^3/\Gamma$
is matched with (\ref{5.40})--(\ref{5.42}) but not with (\ref{2.5}) or (\ref{5.10}).
In any case, $\tilde e^a$ are suitable
one-forms on $\R^4/\Gamma$ which can be used in the ansatz for
instanton solutions.

Let
\beq\label{5.43}
\Acal =\Xt_\m\, \et^\m
\eeq
be an $\su$-invariant connection on the V-bundle $\Ecal$ over
$\R^4/\Gamma$ given in (\ref{3.7}). Here $\et^a$ are given in (\ref{5.37}),
$\et^4:=\diff\tau={\diff r}/{r}$ and $\Xt_\m$ depend only on
$r=\e^\tau$. The explicit form (\ref{5.41}) of $\et^\m$
and the $\Gamma$-action (\ref{3.2}) imply
$\Gamma$-equivariance conditions for the components $\Xt_\m$ given by
\beq \label{5.44}
\g_{_\Gamma}\, \big(\Xt_1+\im\,\Xt_2\big)\,
\g_{_\Gamma}^{-1}=\zeta^{-2}\, \big(\Xt_1+\im\,\Xt_2 \big)\and
\g_{_\Gamma}\, \big(\Xt_3+\im\,\Xt_4\big)\, \g_{_\Gamma}^{-1}=\Xt_3+\im\,\Xt_4\ .
\eeq
For $k\ge 2$ the non-zero blocks of $\Xt_\m$ solving (\ref{5.44}) are given by the matrix elements
\beq \label{5.45}
\big(\Xt_1+\im\,\Xt_2 \big)^{\ell, \ell+2} \ \in \ \Hom
\big(\C^{N_{\ell+2}}\,,\, \C^{N_{\ell}} \big) \and
\big(\Xt_3+\im\,\Xt_4 \big)^{\ell, \ell} \ \in \ \End
\big(\C^{N_{\ell}} \big)
\eeq
for $\ell =0,1,\dots,k$, together with corresponding non-zero blocks
of $\Xt_1-\im\,\Xt_2=-(\Xt_1+\im\,\Xt_2)^\+$ and $\Xt_3-\im\,\Xt_4=-
(\Xt_3+\im\,\Xt_4)^\+$.

In the following we consider only the case of even rank $k=2q$, since
the odd case $k=2q{+}1$ can be reduced to a ``doubling'' of the even
 case. Using the property $\xi^{2q+1}=1$, one has
\beq\label{5.46}
\diag \big(1,\zeta^2, \dots,\zeta^{2k}\big)=\diag\big(1,\zeta^2,
\dots ,\zeta^{2q}, \zeta ,\zeta^3, \dots, \zeta^{2q-1} \big)\ .
\eeq
Then by using the matrix
\beq\label{5.47}
\g_{_\Gamma}=\diag\big(\one_{N_0} ,
\zeta^2\, \one_{N_1},\dots,
\zeta^{2q}\, \one_{N_q},
\zeta\, \one_{N_{q+1}} ,\zeta^3\, \one_{N_{q+2}}, \dots,\zeta^{2q-1}\,
\one_{N_{2q}}\big)
\eeq
in (\ref{5.44}), we obtain the solution
\beq \label{5.48}
\Xt_1+\im\,\Xt_2=\begin{pmatrix}
   0&\phi_1&0&\cdots&0\\0&0&\phi_2&\ddots&\vdots\\ \vdots&\vdots&\ddots&\ddots&0\\0&0&\cdots&0&\phi_k \\
   \phi_{k+1}&0&\cdots&0&0 \end{pmatrix}\and
\Xt_3+\im\,\Xt_4=\begin{pmatrix}
   \r_0&0&\cdots&0\\0&\r_1&\ddots&0\\\vdots&\ddots&\ddots&0\\
   0&\cdots&0&\r_k \end{pmatrix}
\eeq
where $\p_{\ell +1}\in\Hom(\C^{N_{\ell+1}}, \C^{N_{\ell}})$ and $\r_\ell\in\End (\C^{N_{\ell}})$.

In the basis $(\et^\m)=(\et^a, \diff\tau)$ the $\su$-invariant
connection $\Acal$ and the curvature $\Fcal$ are given by
\beq \label{5.49}
\Acal =\Xt_\m\, \et^\m=\widehat{\Xt}_\m\, \widehat{\et}\,^{\m}\with\widehat{\Xt}_\m =\sfrac{1}{r}\,\Xt_\m\and
\widehat{\et}\,^{\m} =r \, \et^\m\ ,
\eeq
\beq \label{5.50}
\Fcal =\frac{1}{r^2}\, \Big(\,\frac14\,\ve_{abc}\, [\Xt_b, \Xt_c] -
\Xt_a\, \Big)\, \bar\h^a_{\m\n}\,\diff y^\m\wedge\diff y^\n
+ \Big(\, \frac{\diff\Xt_a}{\diff\tau}+ [\Xt_4, \Xt_a]- 2\Xt_a+
\frac12\,\ve_{abc}\, [\Xt_b, \Xt_c]\, \Big)\, \et^4\wedge\et^a
\eeq
where we used the identity
\beq \label{5.51}
\h^a_{\m\n}\, \et^\m\wedge\et^\n=\frac{1}{r^2}\, \bar\h^a_{\m\n}\,\diff y^\m\wedge\diff y^\n\ .
\eeq
From (\ref{5.50})
it follows that $\Fcal$ is anti-self-dual, $*\Fcal=-\Fcal$, if $\Xt_a$ satisfy the Nahm equations
\beq \label{5.52}
\frac{\diff\Xt_a}{\diff\tau}=2\Xt_a-\frac12\, \ve_{abc}\, [\Xt_b,\Xt_c]-[\Xt_4,\Xt_a]\ .
\eeq
We obtain a solution by choosing $\Xt_4=0$ and taking
\beq \label{5.53}
\Xt_a=\frac{2r^2}{r^2+\La^2}\, I_a \with [I_a,I_b]=\ve_{ab}^c\, I_c\ ,
\eeq
where the $I_a$ are $\su$ generators in the irreducible representation on the space $\C^N$
with $N=N_0+N_1+\cdots+ N_k$ that fits with the $\Gamma$-equivariant
form (\ref{5.48}). For instance, one can work in the Coulomb branch with
$N_\ell =1$ for all $\ell = 0,1,\ldots ,k$ so that $I_a$ embed the
group $\su$ into $\surm(k{+}1)$.
We thus obtain the configuration
\beq \label{5.54}
\Acal=-\frac{2}{r^2+\La^2}\, \bar\h^a_{\m\n}\,I_a\, y^\m\,\diff y^\n \and
\Fcal=-\frac{2\La^2}{\big(r^2+\La^2 \big)^2} \, \bar\h^a_{\m\n}\,I_a\,\diff y^\m\wedge\diff y^\n\ ,
\eeq
which is exactly the BPST instanton extended from $\R^4$ to $\R^4/\Gamma$.
We again have four moduli: the scale parameter $\La$ and the three parameters of global $\su$ rotations.

The 't~Hooft instanton (\ref{5.34}) is gauge equivalent to the BPST instanton
(\ref{5.54}) on the euclidean space $\R^4$. However, this is not so on the orbifold $\R^4/\Gamma$.
For instance, taking $N_\ell =1$ for $\ell=0,1,\ldots,k$, one can obtain only abelian solutions in the 't Hooft ansatz
(\ref{5.17}) while one has irreducible nonabelian BPST instantons (\ref{5.54}).
Of course, one can transform the solution (\ref{5.54}) to a 't~Hooft-type solution in a
singular gauge, but this transformed solution will not be compatible with $\Gamma$-equivariance,
i.e.~it cannot be projected from $\R^4$ to $\R^4/\Gamma$. On the other hand, 't~Hooft-type solutions
are well-defined on V-bundles $\Ecal$ over the orbifold $\R^4/\Gamma$ if the group $\Gamma$
acts trivially on the fibres of $\Ecal$, i.e.~if $\Ecal =\Ecal_0$, $N=N_0$ and $\g_{_\Gamma}=\one_{N_0}$.
The explicit form of such solutions for $N=N_0=2$ can be found e.g. in~\cite{19,23,24}.

\bigskip

\noindent
{\bf Acknowledgements}

\medskip

\noindent
The work of TAI and OL was partially supported by the Heisenberg-Landau program.
The work of OL and ADP was supported in part by the Deutsche Forschungsgemeinschaft under grant LE 838/13.
The work of RJS was partially supported by the Consolidated Grant ST/J000310/1 from the
UK Science and Technology Facilities Council, and by Grant RPG-404 from the Leverhulme Trust.

\end{document}